# Understanding whole-body inter-personal dynamics between two players using neural Granger causality as the explainable AI (XAI)

Ryota Takamido[a]*, Chiharu Suzuki[b], Jun Ota[a], Hiroki Nakamoto[c],

[a]  Research into Artifacts, Center for Engineering (RACE), School of Engineering, The University of Tokyo, 7-3-1 Hongo Bunkyo-ku, Tokyo, 113-8654, Japan

[b]  Department of Sport Science, National Institute of Fitness and Sports in Kanoya, Kanoya, Kagoshima 891-2393, Japan

[c]  Faculty of Physical Education, National Institute of Fitness and Sports in Kanoya, Kanoya, Kagoshima 891-2393, Japan

*Corresponding author: Ryota Takamido

Center for Engineering, School of Engineering, The University of Tokyo

7-3-1 Hongo Bunkyo-ku, Tokyo, 113-8654, Japan

Email: takamido@race.t.u-tokyo.ac.jp

**Abstract**

Background: Simultaneously focusing on intra- and inter-individual body dynamics and elucidating how these affect each other will help understand human inter-personal coordination behavior. However, this association has not been investigated previously owing to difficulties in analyzing complex causal relations among several body components. To address this issue, this study proposes a new analytical framework that attempts to understand the underlying causal structures behind each joint movement of individual baseball players using neural Granger causality (NGC) as the explainable AI.

Methods: In the NGC analysis, causal relationships were defined as the size of the weight parameters of the first layer of a machine-learning model trained to predict the future state of a specific time-series variable. To verify the approach in a practical context, we conducted an experiment with 16 pairs of expert baseball pitchers and batters; input datasets with 27 joint resultant velocity data (joints of 13 pitchers and 14 batters) were generated and used for model training.

Results: NGC analysis revealed significant causal relations among intra- and inter-individual body components such as the batter's hands having a causal effect from the pitcher's throwing arm. Remarkably, although the causality from the batter's body to pitcher's body is much lower than the reverse, it is significantly correlated with batter performance outcomes ($R^2$=0.69).

Conclusions: The above results suggest the effectiveness of NGC analysis for understanding whole-body inter-personal coordination dynamics and that of the AI technique as a new approach for analyzing complex human behavior from a different perspective than conventional techniques.

## 1. Introduction

Interacting with other humans through our body is one of the fundamental human cognitive motor skills that form the basis of living (handing over objects to others and expressing intentions through hand gestures). These interactive behaviors of humans are called inter-personal coordination (IC) [1]. While IC analysis has been conducted with relatively simple movements, such as synchronizing a part of the body with others in the experimental room [2, 3], IC analysis in sports contexts allows us to gain insights into more complex, whole-body, dynamic, and long-sequential interactions and to verify the findings in more practical environments. Therefore, several studies have analyzed how humans interact with others through their bodies in sports contexts and have obtained useful insights into how humans organize complex IC in sports contexts [4-7].

However, the problem with these approaches is a lack of understanding of whole-body dynamics in intra- and inter-individual interactions. Since previous studies aimed to find the "control variable," which defines the complex interaction among players, such as the distance between two persons [6], the human body was represented by a few key variables. However, because our body has a complex structure consisting of a large number of joints, muscles, and bones, movement is also organized by intra-individual interactions among these body components, called the kinetic chain or energy flow [8, 9], which is necessary to understand the detailed mechanism of IC in sports. For example, in tennis service situations, the motion of the service player is organized by the dynamics of sequential interactions from the lower to the upper body [10]; a receiver observes it and tries to understand the intention through the underlying dynamics in the opponent's body movement [11]. Finally, the receiver also starts to generate a series of interactions within their body as a response to the opponent's action [12]. From the perspective of the dynamical system theory, interpersonal synergies are higher-order control systems formed by coupling the movement-system degrees of freedom of two (or more) actors [13]; however, intra-body synergies are stronger than inter-body synergies [14]. Therefore, to understand whole-body IC, as in the above example, we need to simultaneously focus on intra- and inter-individual whole-body dynamics and verify how these dynamics affect each other. No study has been conducted regarding this aspect because of the high complexity and difficulty in analyzing complex sequential (causal) relationships among many body components.

Recent advances in machine-learning techniques have enabled us to adopt different approaches to understanding the cognitive-motor behavior of humans. An explainable AI (XAI) [15] is a concept that attempts to uncover the black box between the input and output of the machine-learning model and increases the interpretability and explainability of the outcome, such as identifying the reason for the disease diagnosis [16]. Compared to conventional statistical models, one of the advantages of this approach is its high ability to model the nonlinear relationships among a large amount of time-series data. Some recent studies have attempted to identify the characteristics of expert behavior from multiple time-series data [17, 18] and construct an automatic feedback system for novices [19, 20]. Therefore, there is a potential that XAI analysis can help us understand the whole-body IC.

Based on the above background, this study proposes a new approach to understanding the underlying causal relations of whole-body IC using neural Granger causality (NGC) [21] as the XAI. NGC is a state-of-the-art causal estimation technique that identifies nonlinear Granger causal relations among multiple time-series data by referencing the weights of the neural network model trained to predict future states of the target variable based on the previous data of other variables, including the variable itself. It is expected that using NGC, we can clarify causal relations among both intra- and inter-individual joint movements and understand the dynamics of whole-body IC in sports through the "causal graph." Our approach reframes whole-body human movement as information (causality) flow among intra- and inter-individual body joints and analyzes both in the same plane (Figure 1). If we quantitatively measure the causal relations inside and outside the body using NGC, we can apply conventional statistical analysis to identify the key player or joint with a large causality with other players.

Among the many motor skills, the interaction between pitchers and batters in bat-and-ball sports (e.g., baseball or cricket) was set as the target skill in this study for verifying the effectiveness of the approach. This selection is because both players' movements (i.e., pitching and hitting) consist of whole-body movements with a long sequential kinetic chain from the lower to the upper joints [22, 23]. Hence, they should show strong intra-body causal relationships if the NGC analysis can capture these causal structures. Moreover, evidence suggests that although the hitting movement is performed on the flying ball, it is strongly affected by the pitcher's movement before ball release [24-26]. Hence, we can confirm its ability to detect interbody causal relationships. Furthermore, because there is a time lag between the pitching and hitting movements owing to the waiting time for the ball to arrive at the batter [27], we can also confirm whether the NGC can identify such a specific lagged structure behind the motion of both players. Finally, since the hitting movement is considered one of the most difficult skills in the world [28], if we can find a clue for defining the success and failure of hitting through the NGC analysis, it is also worthwhile from a practical viewpoint.

Therefore, the IC between pitchers and batters in bat-and-ball sports is a good example for verifying the effectiveness of our proposed approach in analyzing complex whole-body IC in a sports context. This study aimed to address the following three challenges that are difficult to verify using existing techniques.

*Challenge 1: Quantitative evaluation of the impact of one player's actions on other players.*
The degree to which the information of a player's whole-body movement contributes to the generation of that of other players and how this affects performance outcomes is unclear. Although some existing techniques, such as cross-correlation [29] or transfer entropy [30], can partially address this problem, they have difficulty dealing with a large number of causal relations (27 × 27 joint combinations) that do not ensure linearity and stationarity.

*Challenge 2: Identification of the key body components that organize the IC.* The results of previous research [31, 32] suggest that a key body component contributes to organizing the IC (e.g., the pitcher's throwing arm); however, it is difficult to quantitatively evaluate the contribution of such local information to others' whole-body dynamics in the existing methodologies.

*Challenge 3: Identification of underlying lagged structure inside the IC.* Although there is a time lag in intra- and inter-personal coordination (time lag for kinematic chains and responses to other people), identifying and evaluating the lagged structures in the same plane is difficult. Because sports skills impose strict temporal constraints on players [33], this aspect may also be a critical factor in understanding the mechanism of IC.

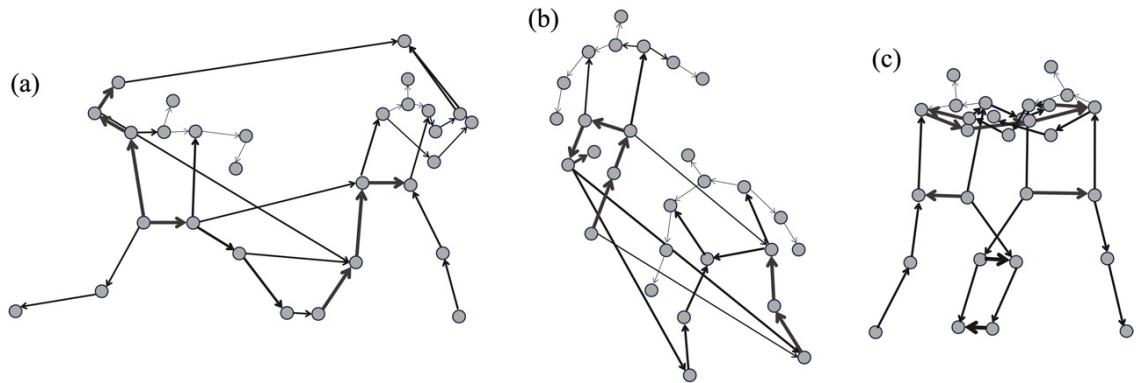

Figure 1. Schematic image of the causality graph for understanding underlying causal relations in whole-body IC of (a) baseball, (b) soccer, and (c) contact sports such as judo. The allow represents the causal direction and size between body joints.

## 2. Methods

This section provides detailed information regarding the methodology of our approach. Note that since there are many "hyperparameters" included in the process of our approach, such as the selection of the feature variables to the network or dawn-sampling rate, we discussed in detail how to tune those parameters and the effect of changing them on the results of this study in the supporting information file. All datasets and sample codes for our analysis are freely available on GitHub (https://github.com/takamido/NGC_data_baseball).

### 2.1 Participants

In total, 23 male expert baseball players, 7 pitchers (6 right-handed, 1 left-handed) and 16 batters (9 right-handed, 7 left-handed), aged 19–23 years (height: 1.73 ± 0.06 m, weight: 75.0 ± 7.9 kg, body mass index: 24.9 ± 1.7 kg/m$^2$) participated in the experiment. The average years of experience in baseball was 12.8 ± 1.9 years. All the participants were on college baseball teams and participated in regular baseball training and competitions. This study was approved by the University's Institutional Review Board for Human Subjects Research and conformed to the principles of the Declaration of Helsinki. All the participants involved in this study provided written informed consent.

### 2.2 Apparatus and procedure

Before starting the experiment, each of the 16 batters was assigned to play against one of the seven pitchers; of the seven pitchers, two pitched against three batters, and the remaining five pitched against two batters. Consequently, 16 pairs of pitchers and batters were selected for this experiment. Because the NGC analysis extracts the relative relationship between two players, replacing one player qualitatively changes the type of interaction between them. Hence, the 16 pairs were treated independently if one player differed.

In the experiment, one pitcher and one batter were instructed to play against each other, as in a real game, until the batter completed 10 swings (including non-contact cases). To decrease the number of pitches and the risk of injury, we instructed the pitchers to pitch the fastball to the center of the strike zone, and the batters were instructed to hit all the balls going there. The mean ball speed of all pitchers was 76.8 mph. When the pitched ball did not pass through the strike zone, the batters could not swing. No other experimental controls were used to obtain data on pitcher-batter interactions in a game-like context.

For more detailed procedures, after receiving the instructions, the pitchers and batters warmed up, as in a regular game, and the reflective markers were set up. The reflective markers were attached to the regio parietalis (head), both sides of the acromion (shoulder), lateral epicondyle of the humerus (elbow), radial styloid process (wrist), greater trochanter of the femur (hip), lateral condyle of the femur (knee), heel, and the top of the shoes. For the batters, reflective markers were also placed on the bat tip and knob. To unify the sense of each body joint between right- and left-handed players, the right-side joints of a right-handed pitcher and batter are described as "back-(joint name)," and the left side body joints are described as "front-(joint name)," and vice versa for left-handed players. The measurements started after the markers were in place. When the participants performed the pitching/hitting movement, the position data of the reflective markers attached to each batter and pitcher's body were measured at 250 Hz using 32 synchronized optical motion capture cameras (16 Raptor-E and 16 Kestrel2200 cameras, Motion Analysis Corp, Santa Rosa, USA). In addition, the performance of pitchers, such as pitched ball velocity, was measured using a trackman system

(TRACKMAN, Vedbaek, Denmark), and the hitting results, such as contact/non-contact and in-field/out-field, were manually recorded for analysis.

### 2.3 Whole-body IC analysis using NGC

#### 2.3.1 Dataset generation for NGC analysis

Based on the collected motion-capture data of the pitcher and batter, the input datasets for NGC analysis were generated as follows: First, a low-pass filter with a cutoff frequency of 10 Hz was applied to the motion-capture data, and the resultant velocity of each joint was calculated as the feature value. The whole-body motion of the pitcher and batter was represented by a set of resultant velocities of each joint, and this study attempted to capture the characteristics of "velocity interaction" between two players. A total of 27 resultant velocities of each joint, comprising 13 variables for the pitcher's joints (head, shoulders, elbows, wrists, hips, knees, and heels) and 14 variables for the batter's joints and tool (13 joints and bat tips), were calculated. Subsequently, to unify the timescale of each trial, the timing of the pitcher's ball release was identified by detecting the time of the peak value of the back-wrist velocity of the pitcher, and the data were clipped from 2 s before and 0.5 s after ball release. Finally, the resultant velocity of each joint was normalized to a range of 0 to 1 to equalize the effect of increasing or decreasing the weights of the neural network and downsampled from 250 Hz to 50 Hz to capture the interactions on a long-term scale. Figure 2 shows an example of the standardized resultant velocity of each joint measured in the experiment. This study attempted to understand the underlying causal structure among these time-series data through NGC analysis while removing the boundary of each body. As a result of these processes, an input dataset of 125 points (50 Hz × 2.5 s) × 27 variables × 10 swings was generated for all 16 pairs.

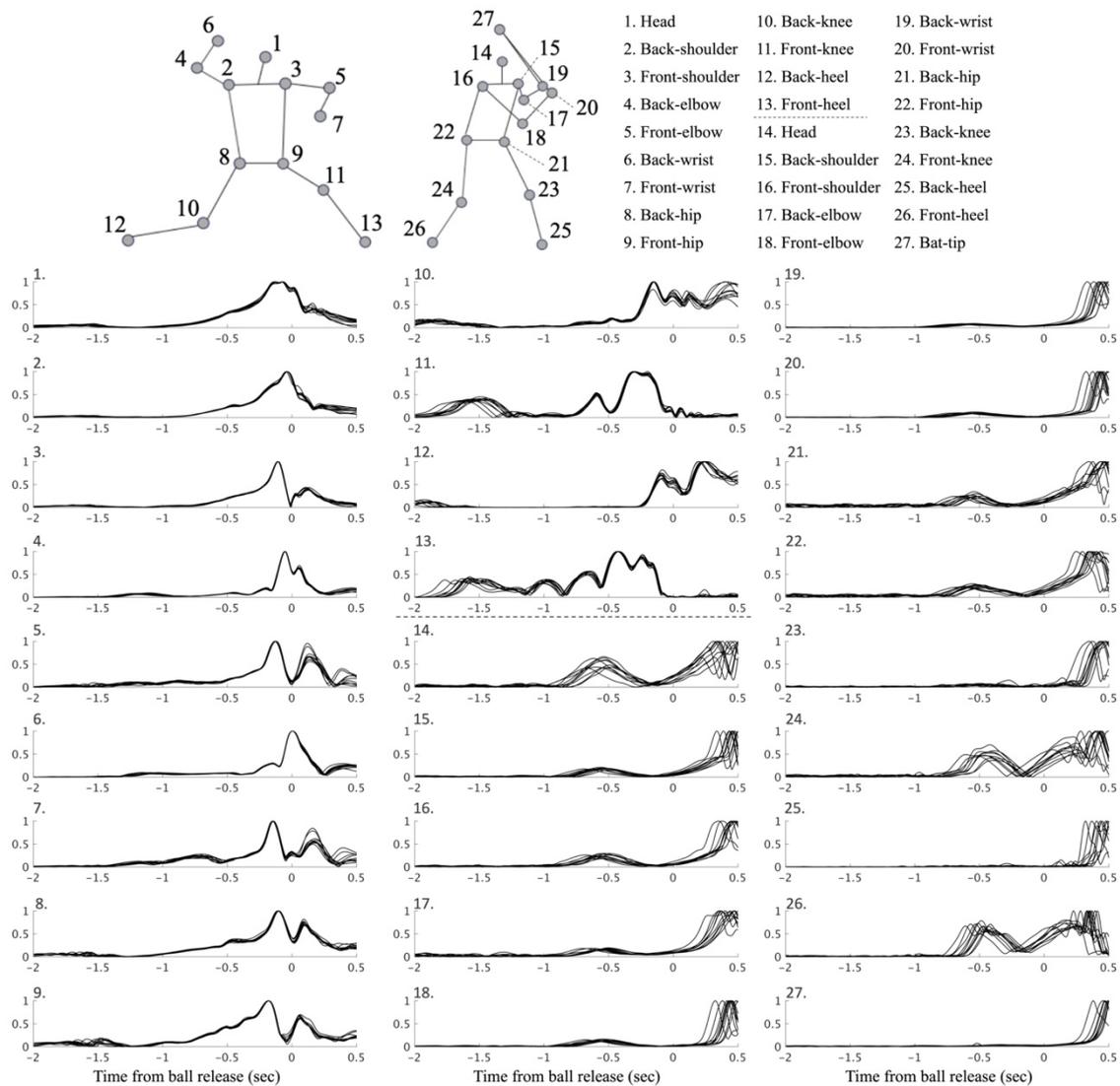

Figure 2. Example of input feature variables, i.e., normalized resultant velocity data of 27 joints of a pitcher (1-13) and a batter (14-27) clipped from 2 s before and 0.5 s after ball release (10 pitches for one pair of pitcher and batter).

**2.3.2 NGC calculation from generated input datasets**

To understand the underlying causal relationships in the input datasets, as shown in Figure 2, we calculated the NGC for each of the 16 pairs of datasets. The NGC used in this study was calculated using a component-wise multilayer perceptron (cMLP) proposed by Tank et al. [21]. NGC calculations using our datasets are summarized below.

Figure 3 shows a schematic of the cMLP used in this study for NGC calculation. The cMLP is trained to predict the velocity of a specific joint at a specific time $x_{i,t}$ based on the past velocity of the joints of the pitcher and the batter (including the target joint itself), whereas the available range for past data is set by a maximum lag $K$. Because the weight of the non-causal parameter to $x_{i,t}$ converges to zero during training and the size of the weights represents the strength of the Granger causality, we can identify the causal relationship among each joint movement of the pitcher's and the batter's bodies by referencing the weight parameters of the trained model. If the batter's movement is influenced by the pitcher's movement, the weight parameters of the pitcher's joints increase when predicting the batter's movement.

Formally, given MLP with $L-1$ layers, the vector $h_t^l \in R^H$ represents the values of the $m$-dimensional $l$th hidden layer at time $t$, weights parameter $\boldsymbol{W} = \{W^1, \ldots, W^L\}$, and weights at the first layer across time lag $K$, $W^1 = \{W^{11}, \ldots, W^{1L}\}$, and the bias parameter for each layer $\boldsymbol{b} = \{b^1, \ldots, b^L\}$, the output from the first hidden layers, is calculated as follows:

$$h_t^1 = \sigma\left(\sum_{k=1}^{K} W^{1k} \boldsymbol{x}_{t-k} + b^1\right), \tag{1}$$

where $\sigma$ is the activation function. After passing through the $L-1$ hidden layers, the time-series output $x_{ti}$, is given by a linear combination of units in the final hidden layer:

$$x_{ti} = g_i(x_{<t}) + e_{ti} = W^L h_t^{L-1} + b^L + e_{ti}, \tag{2}$$

where $e_{ti}$ is modeled as mean-zero Gaussian noise, $x_{<ti} = (\ldots, x_{(t-2)i}, x_{(t-1)i})$, and $g_i$ is a nonlinear function. The NGC is defined by the weight parameter in Equation (2), which was adjusted through the training process using the following penalty equation:

$$\min_{w} \sum_{t=K}^{T} \left(x_{i,t} - g_i\left(x_{(t-1):(t-K)}\right)\right)^2 + \lambda \sum_{j=1}^{p} \Omega(W_{:j}^1), \quad (3)$$

where $\Omega(W_{:j}^1)$ is a regularization term that applies the penalty to the sum of the weights at the first layer for input time series data and helps to avoid overestimating the number of causal relations. Similar to the original article [21], this study adopts a *group sparse group lasso* penalty. Using Equation (3), the weights of the non-causal parameter converge to zero.

Specifically, this study adopted a cMLP with one hidden layer and 32 hidden units for NGC calculation; the maximum time lag $K$ was set as 1.0 s (50 points), and a rectified linear unit was used for the activation function. During model training with the generated datasets, the model was separately trained on the datasets of each pair. Because the data from the 10 swings were batched, the network was trained to identify their common features. The iterative shrinkage thresholding algorithm was used for training; the learning rate was 0.05, regularization parameter $\lambda$ was 0.003, and iteration number was 2000. The implementation was performed on an NVIDIA V100 GPU with Tensor Core and exhibited high prediction accuracy ($R^2 > 0.9$). The training time was less than 2 min for all pairs. The effect of hyperparameters for training, such as $\lambda$, is also verified in the supporting information file.

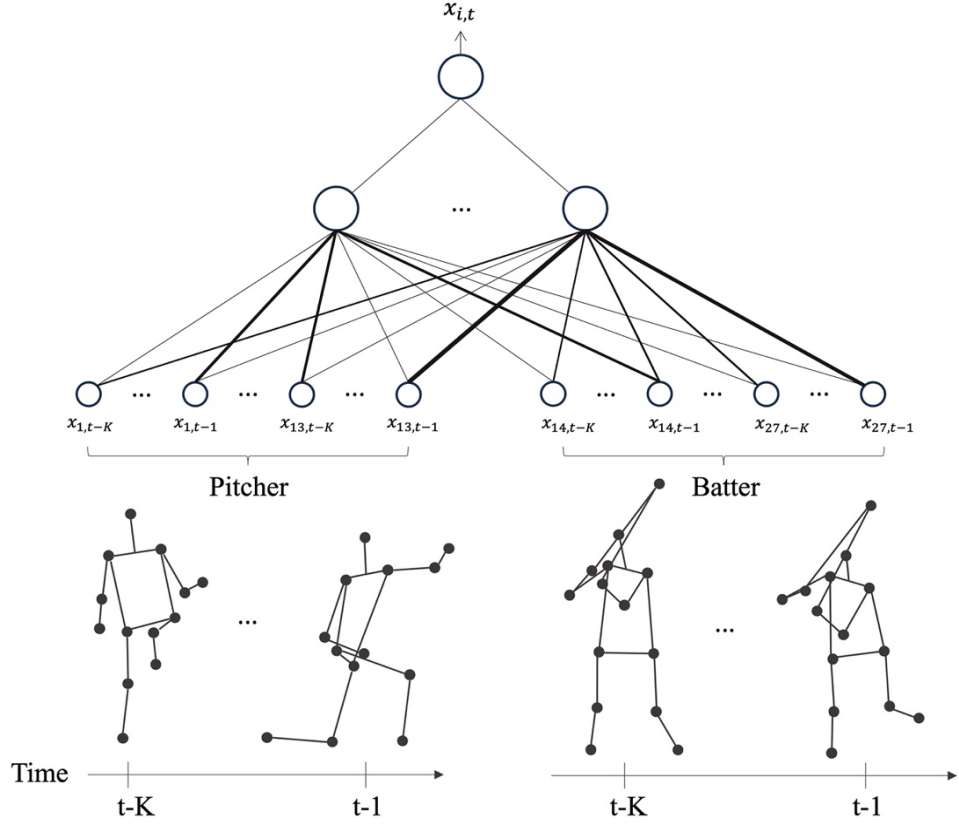

Figure 3. Schematic image of cMLP models for NGC calculation based on the motion-capture data of a baseball pitcher and batter. This model predicts the velocity of a specific joint at a specific time $x_{i,t}$ based on the past velocity of the joints of the pitcher and batter (including the target joint itself) while the available range for past data is set at a maximum lag $K$. The weights of the first layer correspond to the strength of the NGC, and darker colored lines indicate larger weights.

**2.3.3 Analysis of obtained NGC parameters**

Because of the training process, the whole-body IC between the pitcher and the batter was quantified using NGC $ngc_{i,j,k}$, which represents the strength of the causal relationship from joint $j$ to $i$ with a specific time lag $k$. Here, we describe how to address the three challenges mentioned in the Introduction through statistical analysis of the obtained NGCs.

*Analysis 1: Quantitative evaluation of the impact of one player's actions on other players*

To address the first challenge, we quantified the size of the interaction between the intra- and inter-individual bodies of pitchers and batters using the following procedure. First, the NGC matrix $C_{NG} =$

$\{ngc_{1,1}, ngc_{1,2}, \ldots, ngc_{1,27}, \ldots, ngc_{27,27}\}$ was calculated for each pair by taking the sum of causality over the allowed time lag.

$$ngc_{i,j} = \sum_{k=1}^{K} ngc_{i,j,k}, \quad (4),$$

where $ngc_{i,j}$ indicates the size of the NGC from joints $j$ to $i$ (i.e., whether the motion of joint $j$ affects that of joint $i$). Figure 4 shows a schematic image of the NGC matrix. The upper left and lower right regions indicate intra-personal causality, that is, NGC from the pitcher's joints to the pitcher's joints and from the batter's joints to the batter's joints, respectively. The upper right and lower left regions indicate inter-personal (inter-body) causality, corresponding to that from the pitcher's joints to the batter's joints and the batter's joints to the pitcher's joints, respectively. Therefore, we can quantify the mean strength of both the intra- and interbody causal relationships using the following equations:

$$ngc_{pp} = \frac{1}{13*(13-1)} \sum_{\substack{i=1 \\ i \neq j}}^{13} \sum_{j=1}^{13} ngc_{i,j}, \quad (5)$$

$$ngc_{bb} = \frac{1}{14*(14-1)} \sum_{\substack{i=14 \\ i \neq j}}^{27} \sum_{j=14}^{27} ngc_{i,j}, \quad (6)$$

$$ngc_{pb} = \frac{1}{14*13} \sum_{i=14}^{27} \sum_{j=1}^{13} ngc_{i,j}, \quad (7)$$

$$ngc_{bp} = \frac{1}{13*14} \sum_{i=1}^{13} \sum_{j=14}^{27} ngc_{i,j}, \quad (8)$$

where $ngc_{pp}$ and $ngc_{bb}$ indicate the mean Granger causality among the pitcher and batter's body joints and $ngc_{pb}$ and $ngc_{bp}$ indicate the mean Granger causality from the pitcher to the batter's body and the batter to the pitcher's body, respectively. To focus on the causal relations between different body joints, we removed the diagonal component of $C_{NG}$ that represents causality from the same joint (autoregression). Because the penalty term in Equation (3) forces the sum of the overall weights to decrease, we can evaluate and compare the stability of individual movement patterns and the degree of influence of other players in the same plane. If individual movement patterns are highly stable and less influenced by others, $ngc_{pp}$ and $ngc_{bb}$ are predicted to be significantly larger than $ngc_{pb}$ and $ngc_{bp}$.

To statistically test whether there were significant differences in the size of mean causality among intra- and inter-individual body joints. First, the mean values of four causal indexes ($ngc_{pp}$, $ngc_{bb}$, $ngc_{pb}$, $ngc_{bp}$) were calculated among the 16 pairs. Then, a one-way repeated-measures ANOVA was conducted, and a post-hoc (Holm–Bonferroni) pairwise comparison was used to determine whether there was a significant difference in the mean value of the four causal indexes. Furthermore, to verify whether the size of the indexes was related to the batter's performance outcome, the mean contact rate and in-field rate were also calculated as the performance indexes for each pair based on the records in the experiment, and a one-way MANOVA was conducted for the contact rate and in-field rate while considering the four indexes of IC and mean ball speed of each pitcher as dependent variables. If significance was observed, a post-hoc ANOVA was conducted to identify the index that significantly affected the performance outcomes.

These analyses were conducted using R Statistical Software (R Development Core Team, 2016), and the significance level was set at 0.05. The effect sizes were assessed using the partial eta-squared ($\eta^2$) and Cohen's *d* for post-hoc ANOVA and *t*-tests. Furthermore, because estimating the statistical power before conducting this analysis was difficult, we performed a post-power analysis to calculate the smallest possible effect size that could be detected with our sample (16 pairs) using G Power software [34]. We decided that an alpha error probability would determine a statistical significance of <0.05, and the power was set to >0.80. The results indicated that the smallest possible effect sizes $\eta^2$ and *d* were 0.21 and 0.75.

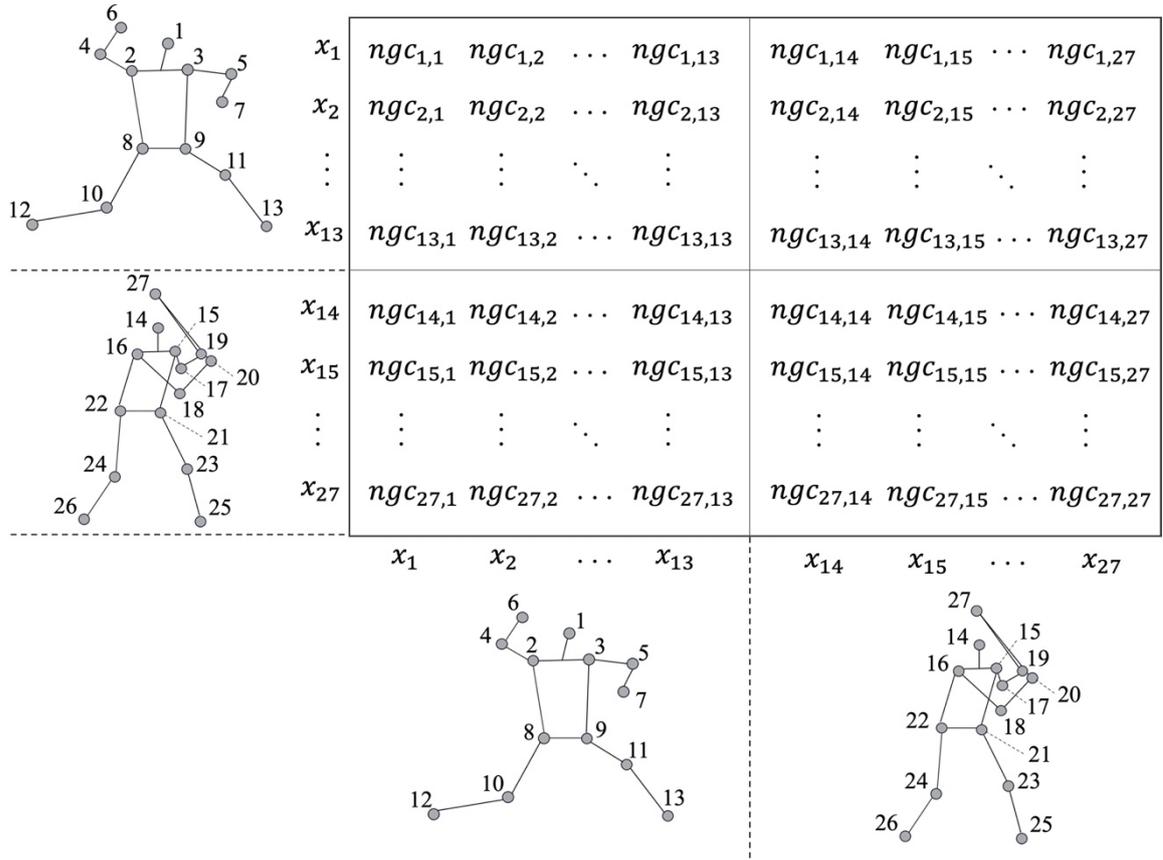

Figure 4. NGC matrix obtained from cMLP training. $ngc_{i,j}$ indicates whether the motion of joint $j$ is the Granger causality of the motion of joint $i$.

*Analysis 2: Identification of the key body components that organize the IC.*

To identify the key body component of the IC, we calculated the mean values of $ngc_{i,j}$ among the 16 pairs, $\overline{ngc_{i,j}}$. The specific components of $\overline{ngc_{i,j}}$ represent inter-personal causality (lower left and upper right areas of Figure 3).

$$\{\overline{ngc'_{i,j}}\} = \{\overline{ngc_{i,j}}\}_{i=14:27, j=1:13}, \tag{9}$$

$$\{\overline{ngc''_{i,j}}\} = \{\overline{ngc_{i,j}}\}_{i=1:13, j=14:27}, \tag{10}$$

where $\{\overline{ngc'_{i,j}}\}$ indicates the set of NGCs from a specific pitcher's joints to a specific batter's joints,

and $\{\overline{ngc''_{i,j}}\}$ indicates that of the batter to the pitcher. If its component shows a positive value, it indicates an NGC from joint $j$ to joint $i$ (e.g., the pitcher's back hand to the batter's front wrist). Since NGC is the non-negative variable ($\overline{ngc_{i,j}} \geq 0$ for all $i, j$), we set the threshold of the mean value of $C_{NG}$ excluding diagonal components to capture the too-small causality in the statistical test.

To statistically verify whether a joint has a significantly large NGC compared to other player's joints, we conducted one-sample *t*-tests with the above threshold for each component of $\{\overline{ngc'_{i,j}}\}$ and $\{\overline{ngc''_{i,j}}\}$. Benjamini & Hochberg's (BH) method was applied to adjust significance, and the other settings for the statistical analysis are the same as in Analysis 1. The smallest possible effect *d* was calculated as 0.65 for one-sample *t*-tests to detect one-tailed significance.

*Analysis 3: Specifying underlying lagged structure inside the whole-body IC.*

Finally, to address the challenge of quantifying the lagged relations, the components of the NGC matrix $ngc_{i,j}$ were replaced by $l_{ij}$, which indicates the time lag that maximizes the NGC from joint $j$ to $i$:

$$l_{ij} = \operatorname*{argmax}_{1 \leq k \leq K}(ngc_{i,j,k})/50 \qquad (11)$$

Same as in Analysis 1, the mean time lags in both intra- and inter-body coordination, $l_{pp}$, $l_{bb}$, $l_{pb}$, $l_{bp}$ were calculated by replacing the $ngc_{i,j}$ in Equations (5)–(8) with $l_{ij}$. These four lag indexes were calculated for each participant, and the same statistical analysis as in Analysis 1 was applied. To avoid including the data of an NGC that was too small, the same threshold as in Analysis 2 was applied to calculate the four lag indexes.

**Results**

Figure 5 shows an example of the NGC matrix $C_{NG}$ obtained from the weights of the trained model and the causal graph between the pitcher and the batter. As shown in this figure, through NGC analysis, we can understand the characteristics of the whole-body IC between the pitcher and batter (e.g., the causality from the pitcher to the batter is larger than that from the batter to the pitcher, and the batter tends to have causality from the pitcher's throwing arm). The detailed results of each analysis are presented below.

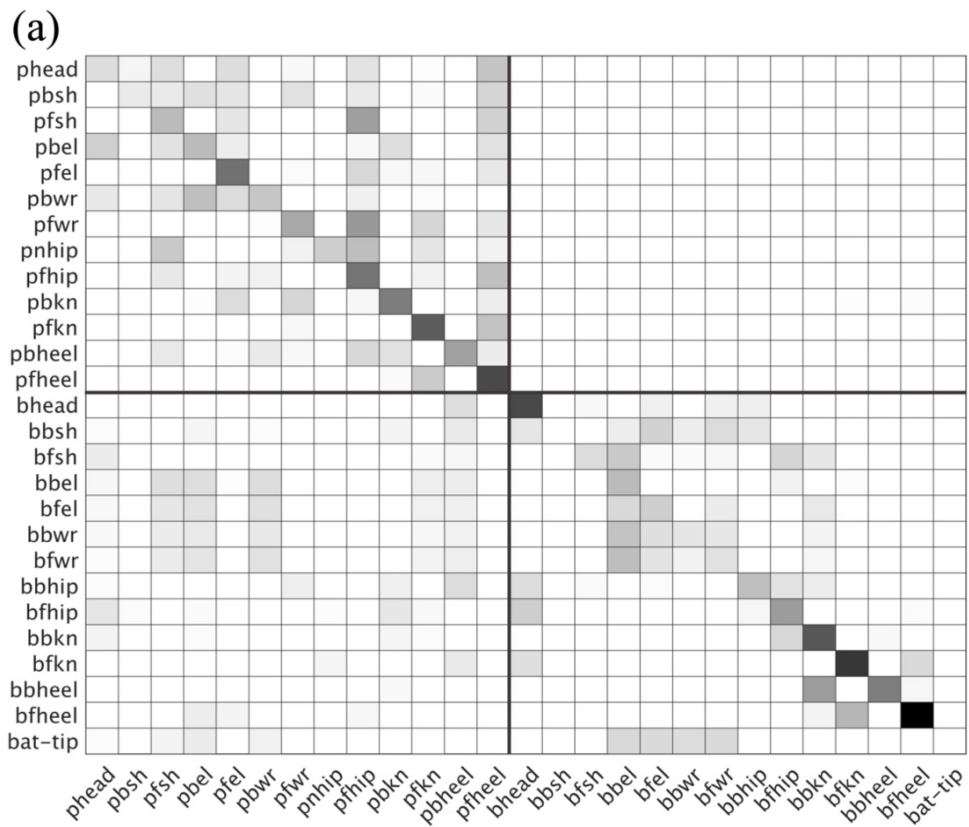

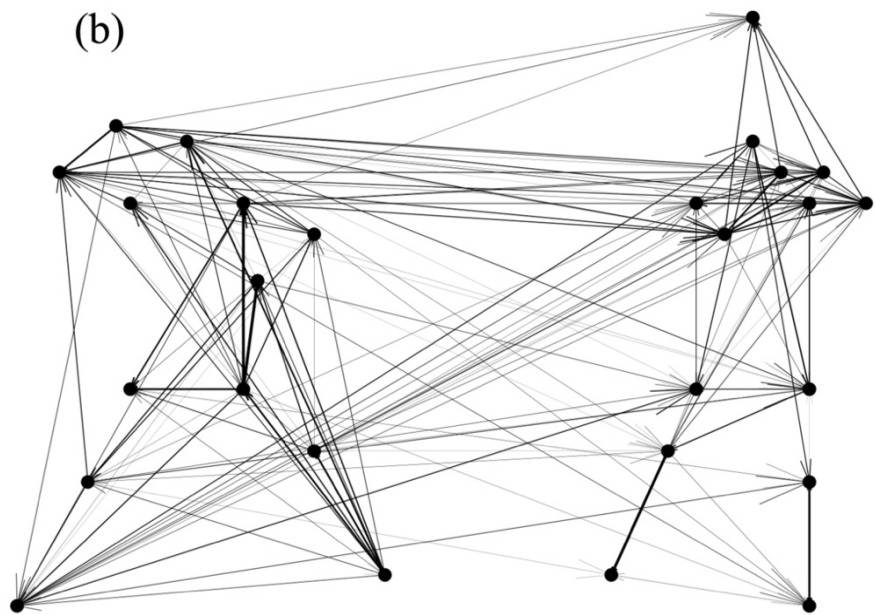

Figure 5. (a) Example of the NGC matrix obtained from the analysis (one pair of pitcher and batter). The terms "pb_" and "pf_" represent the pitcher's back and front joints (right and left for right-handed), and "bb_" and "bf_" represent the batter's back and front joints. The color intensity indicates the degree of causality. (b) Causal graph representation of the matrix. The arrow and width of lines indicates the direction and degree of NGC, respectively.

*Analysis 1: Quantitative evaluation of the impact of one player's actions on other players*

Figure 6 shows the mean and standard deviation of the four causality indexes ($ngc_{pp}$, $ngc_{bb}$, $ngc_{pb}$, $ngc_{bp}$) among 16 pairs of data. A one-way repeated measure ANOVA suggested significant differences among four indexes ($F(3, 60) = 2.75$, $p < .01$, $\eta^2 = 0.90$). Post-hoc pairwise comparisons revealed that the mean NGC from a pitcher's body to another pitcher's body ($ngc_{pp}$) is significantly larger than that among batter's joints ($ngc_{bb}$), from the pitcher to the batter's joint ($ngc_{pb}$), and from the batter to the pitcher's joint ($ngc_{bp}$), with higher statistical power than the criteria ($t(15) = 4.1$, $p < .01$, $d = 2.9$, $t(15) = 13.2$, $p < .01$, $d = 4.7$, $t(15) = 59.0$, $p < .01$, $d = 20.9$, respectively). $ngc_{bb}$ is significantly larger than the $ngc_{pb}$ and $ngc_{bp}$ ($t = 8.3$, $p<.01$, $d = 2.9$, $t(15) = 14.7$, $p < .01$, $d = 5.2$, respectively), and $ngc_{pb}$ is significantly larger than the $ngc_{bp}$ ($t(15) = 8.4$, $p < .01$, $d = 3.0$). Therefore, while the pitcher has a high causality within his body and few causalities from the batter's motion, the batter has more causal relationships with the pitcher's motion, and its motion is highly constrained; hence, they have an asymmetrical relationship.

Moreover, the MANOVA analysis suggested a significant effect of the four NGC indexes on the in-field rate ($F(1, 14) = 10.9$, $p < .01$, $\eta^2 = 0.60$), with no significant effect on the contact rate ($p > .05$). The post-hoc analysis by a separate ANOVA indicated that the factor of ball speed, $ngc_{pp}$ and $ngc_{bp}$, has significant effects with a large statistical power on the mean in-field rate among 10 swings of the batter ($F(1, 14) = 6.6$, $p=.02$, $\eta^2 = 0.32$, $F(1, 14) = 9.4$, $p < .01$, $\eta^2 = 0.40$, $F(1, 14) = 32.1$, $p < .01$, $\eta^2 = 0.69$, respectively). Therefore, as shown in Figure 7, although the mean ball speed of each pitcher also significantly affects the hitting result, the causal relationship between the pitcher and the batter has a larger effect. Although the size of the NGC from the pitcher's joints to the batter's joints $ngc_{bp}$ is small, whether the hitting ball falls into the in- or out-field depends highly on it.

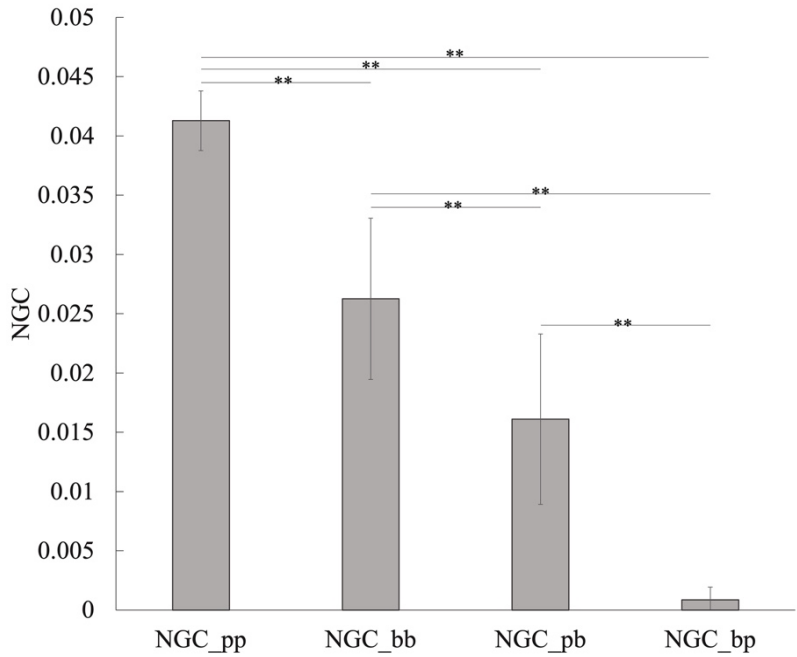

Figure 6. The differences in mean NGC from the pitcher's body to the pitcher's body (NGC_pp), the batter's body to the batter's body (NGC_bb), the pitcher's body to the batter's body (NGC_pb), and the batter's body to the pitcher's body (NGC_bp). The error bar indicates the standard distribution

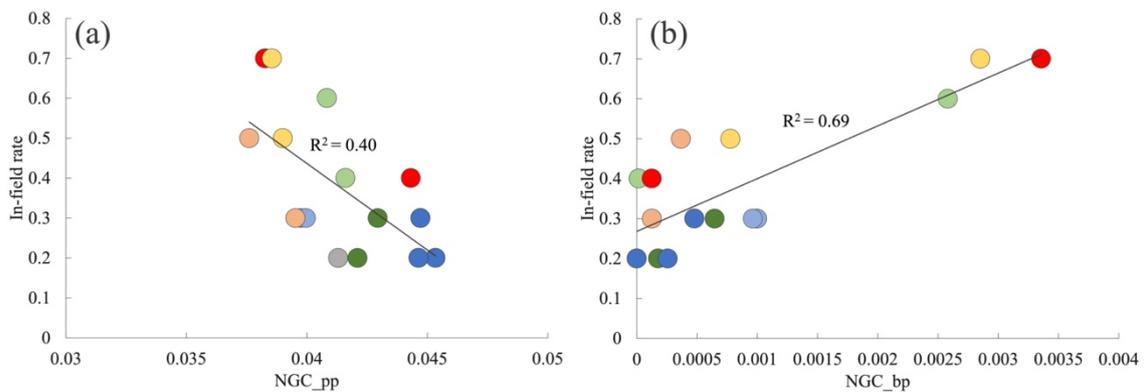

Figure 7. The relationship between NGC and performance outcomes. (a) $ngc_{pp}$ and in-field rate of 10 pitches (b) $ngc_{bp}$ and in-field rate. Colors are classified based on the differences in the opponent pitcher.

*Analysis 2: Identification of the key body components that organize the IC.*

Table 1 presents the results of Analysis 2. A total of 10 pitcher joints had a significantly larger NGC than the threshold (0.015), and six joints had a larger effect size than the criteria but did not show significance. Figure 8 shows a graph of the NGC in Table 1. From Table 1 and Figure 8, we can identify the key body components of the pitcher that organize the batter's motion, such that the pitcher's throwing arm (back elbow and wrist) has a high causality with the batter's hand motion (front and back wrist, bat tips). In addition, no batter joints had high causality to the pitcher's joints ($p > .05$).

Table 1. The summary of the results of Analysis 2. The NGC in this table indicates the size from the pitcher's joint to the batter's joint.

| Pitcher's joint | Batter's joint | mean NGC | SD | adjusted $p$ | $t(15)$ | Cohen's $d$ |
|---|---|---|---|---|---|---|
| back-elbow | front-wrist | 0.08 | 0.05 | * | 4.87 | 1.16 |
| back-wrist | back-wrist | 0.06 | 0.04 | * | 4.48 | 1.06 |
| back-elbow | back-wrist | 0.07 | 0.05 | * | 4.45 | 1.06 |
| back-elbow | bat-tip | 0.06 | 0.04 | * | 4.39 | 1.04 |
| back-wrist | back-wrist | 0.06 | 0.04 | * | 4.34 | 1.03 |
| back-wrist | front-wrist | 0.06 | 0.04 | * | 4.32 | 1.03 |
| back-elbow | front-elbow | 0.06 | 0.04 | * | 4.29 | 1.02 |
| back-elbow | back-elbow | 0.06 | 0.04 | * | 4.15 | 0.99 |
| back-wrist | bat-tip | 0.05 | 0.04 | * | 4.15 | 0.98 |
| back-wrist | front-elbow | 0.05 | 0.04 | * | 3.76 | 0.89 |
| back-knee | back-elbow | 0.06 | 0.05 | n.s. | 3.32 | 0.79 |
| back-elbow | back-knee | 0.06 | 0.05 | n.s. | 3.29 | 0.78 |
| back-heel | back-shoulder | 0.04 | 0.03 | n.s. | 3.08 | 0.73 |
| back-knee | front-hip | 0.05 | 0.04 | n.s. | 3.04 | 0.72 |
| back-knee | back-wrist | 0.05 | 0.04 | n.s. | 3.00 | 0.71 |
| back-wrist | back-knee | 0.05 | 0.05 | n.s. | 2.92 | 0.69 |

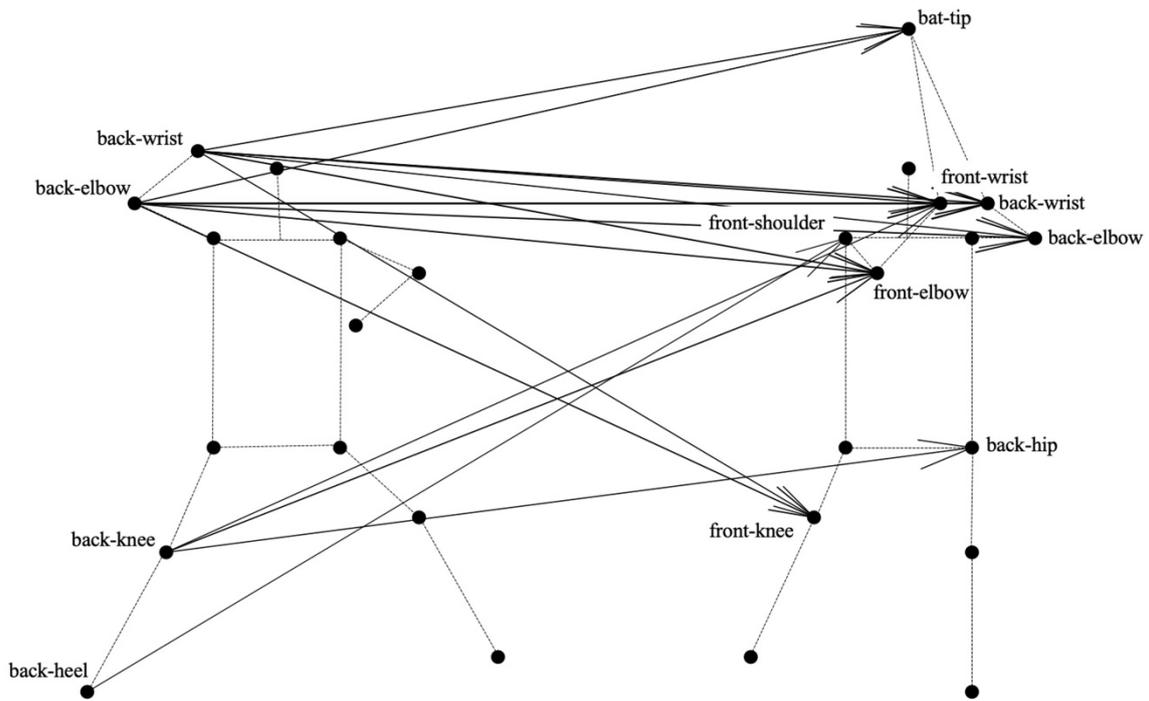

Figure 8. The causal graph of the NGC in Table 1.

*Analysis 3: Specifying underlying lagged structure inside the whole-body IC.*

Figure 9 (a) presents the mean and standard deviation of the three time-lag indexes ($l_{pp}$: 0.13±0.01 s, $l_{bb}$: 0.07±0.04 s, $l_{pb}$: 0.49±0.06 s). Since more than 90% of combinations of joints used for the calculation of the $l_{bp}$ showed a smaller NGC than the threshold for all pairs, we excluded the variable from the analysis. A one-way repeated measure ANOVA revealed significant differences among four indexes ($F(2, 45) = 3.20$, $p < .01$, $\eta^2 = 0.95$). Post-hoc pairwise comparisons revealed that the mean time lag from the pitcher's joints to the batter's joints ($l_{pb}$) was significantly longer than that among the pitcher's joints ($l_{pp}$) and the batter's joints ($l_{bb}$) ($t(15) = 26.4$, $p < .01$, $d = 6.2$, $t(15) = 5.6$, $p < .01$, $d = 1.4$, respectively), and $l_{pp}$ was significantly longer than $l_{pp}$ ($t = 8.3$, $p<.01$, $d = 2.9$). Therefore, there are significantly different lag relationships within the pitcher's body, the batter's body, and among the pitcher and batter's body. Figure 9 (b) shows the plots of the mean NGC between each joint at each time lag.

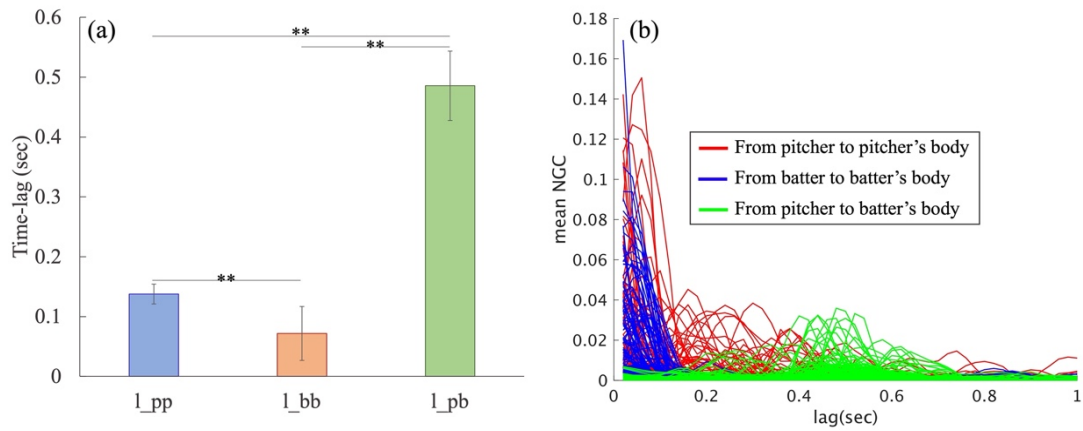

Figure 9. The results of Analysis 3. (a) Mean time lag among the batter's joints (blue), the pitcher's joints (red), and the pitcher's to batter's joints (green) that maximizes the NGC between specific joints. The error bar indicates the standard deviation of the 16 pairs. (b) Mean NGC between each joint at each time lag among the 16 pairs of data. All combinations of joints in each domain are plotted here (14 × 13 lines for blue, 13 × 12 lines for red, and 13 × 14 lines for green).

**Discussion**

Analysis 1 revealed the characteristics of the causal relationship between pitchers and batters. There were significant differences among the four causal indexes ($ngc_{pp}$, $ngc_{bb}$, $ngc_{pb}$, $ngc_{bp}$), and the percentages of the sum of all causalities were approximately 49%, 31%, 19%, and 1%. As discussed in previous research [27], this asymmetric relationship seems to reflect the differences in their roles in the game, that is, the differences in task constraints. While the pitcher is in the dominant position and controls the initiation and end of the pitching movement, the batter must observe and follow it while predicting the ball's trajectory. Although the previous study [27] partially analyzed this asymmetric relationship, the present study first quantified it by considering the causal relationships of all the joints of the two players' bodies. From a simple calculation, approximately 38% of NGC for the batter's movement generation was derived from the pitcher's movement ($ngc_{pb}/(ngc_{bb} + ngc_{pb})$=0.38), suggesting how the batter's movement is constrained by the pitcher's movement.

Furthermore, the results of Analysis 1 also suggested that each $ngc_{pp}$ and $ngc_{bp}$ had negative and positive effects on the performance outcome of the batter, that is, the in-field rate among 10 swings (Figure 7). Remarkably, $ngc_{bp}$ that indicates the mean NGC between the batter's and the pitcher's joints showed a large effect on the performance ($\eta^2 = 0.69$), although it accounts for only a small percentage in the total causalities (approximately 1.0%). These results can be interpreted in several ways. First, the movement stability and independence of the pitcher may be key factors in bat-and-ball sports. Because a small difference in the release parameters, such as the release angle, causes a large difference in the pitched ball [35], a small NGC from the batter to the pitcher's body may slightly change the pitcher's movement and significantly impact the performance outcome. Another possibility is that the batter's ability to predict the pitcher's movements reflects these indexes. If the batter accurately anticipates and models the pitcher's movement and consistently moves in advance of the pitcher throughout the 10 swings, the NGC from the batter's joint to the pitcher's joint increases. Since evidence suggests that expert batters tend to move against the pitcher more quickly than novices [26, 36] and the batter's anticipatory skill significantly affects pitched-ball recognition [37] and impact timing [38-39], the latter seems to have a larger possibility than the former.

In addition, the results of Analysis 2 revealed the key pitcher's body components that have a large causality with the batter's movement. There are two main types of information resources. The first is the motion of the pitcher's throwing arm (back elbow and back wrist). This helps to determine the critical factors for successful hitting, such as ball release timing, ball type, and speed, and the result matches the existing findings [31-32]. Hence, it is natural for the batter to observe and follow the pitcher's throwing arm motion. The second type is the pitcher's lower body (back knee and back heel), which suggests that the batter also adjusts to the pitcher's weight-shift movement. Because the pitcher's weight shift movement occurs before the arm-throwing motion, this result suggests that the batter first uses the pitcher's lower body information to roughly tune his hitting movement and then uses the arm-throwing movement for fine-tuning. The findings of multiple information resources at different movement stages match the existing model of anticipatory behavior in a striking sports context [40], and the present study demonstrates this through NGC analysis.

NGC analysis can also detect the lagged structure in the whole-body IC, as shown in the results of Analysis 3. The differences between the $l_{pp}$ and $l_{bb}$, which represent the mean time lag that maximizes the NGC among intra-pitcher and batter joints, are considered to be caused by the difference in the dynamics between pitching and hitting movements. While the pitcher has sufficient time to form the energy (causal) flow from the lower joints to the upper joints, the batter has temporal constraints to do so with a limited amount of time until the ball arrives. A plot in Figure 9 for $l_{pp}$ shows causal relationships with a large time lag (0.2–0.4 s). Moreover, as discussed in the previous study [27], since the value of the $l_{pb}$ (0.49 s) corresponds to the ball travel time of the pitcher (approximately 0.50 s for 76.8 mph [41]), the batter interacts with the pitcher with a specific time lag.

Summarizing the results of the three analyses and the above discussions, we can understand the dynamics of whole-body IC in bat-and-ball sports as follows: first, the pitcher starts generating a causality flow within his body with a short time lag. Then, the flow is transmitted to the batter's body with a longer time lag that corresponds to the ball travel time, and the batter also starts to generate a causality flow among his body joints while adjusting some parts of his movement based on the pitcher's key joint movement, such as the throwing arm. Although this causality flow is mostly one-way from the pitcher to the batter, the hitting performance will improve if the batter shows the opposite flow.

The insights presented in this study are the first to be revealed by NGC analysis and suggest its effectiveness in understanding whole-body inter-personal coordination dynamics, and more broadly, the effectiveness of the AI technique as a new approach for analyzing complex human behavior from a different perspective than conventional techniques. Although this study focused on the whole-body IC in bat-and-ball sports as the representative example, NGC analysis can be widely applied to human interactive behaviors other than sports activities. Therefore, we expect that application of NGC analysis to various motor skills and contexts will lead to the accumulation of new findings in understanding complex human collective behavior.

This study had some limitations. First, because we mostly focused on the movement interaction between the two players, it is necessary to verify the relationship between the perceptual and cognitive aspects of both players. For example, if the batter changes the attention area from some parts of the pitcher's body to others or part of his own body, it may dramatically change the causal relationship between the two players. Furthermore, although the pitchers only pitched the fastball in the experiment, more variation in ball types may increase the use of the pitcher's advanced kinematic information for the batter and the causality from the pitcher to the batter. Finally, because this analysis can only capture the relative relationships between two players, further control experiments, such as controlling opponents' movement and reaction in a virtual reality environment, are necessary to understand the detailed mechanism of whole-body IC.

**Conclusion**

This study attempted to explain the underlying dynamics of whole-body IC between pitchers and batters using the NGC as the XAI. We obtained several novel findings regarding the size of causality among each intra- and inter-personal body joint in pitchers and batters, its relationship with performance outcome, and the key pitcher's body components that organize the batter's movement and lagged structure inside the whole-body IC. Applying the NGC analysis to whole-body IC in sports can provide several insights that may be difficult to obtain from conventional approaches. Therefore, using this new analytical approach combined with traditional methods, we can understand complex human behavior in sports contexts more comprehensively. Further control experiments, such as those controlling opponents' movement and reaction in a virtual reality environment, are necessary to understand the detailed mechanism of whole-body IC.

**Acknowledgments**

The data used in this study were captured at the Sports Performance Research Center of the National Institute of Fitness and Sports, Kanoya (NIFS SPRC). The authors thank the members of the NIFS SPRC and NIFS for their technical assistance with the experiments.

**Funding**

This work was supported by Japan Society for the Promotion of Science (Grant number: JP22K17712).

**Author Contributions**

Ryota Takamido: Conceptualization, Data curation, Formal analysis, Funding acquisition, Investigation, Roles/Writing - original draft, Visualization

Chiharu Suzuki: Methodology, Resources

Jun Ota: Conceptualization, Supervision

Hiroki Nakamoto: Methodology; Supervision, Project administration

**Supporting information**

**Aim of this Supporting Information file**

As is widely known, the performance of the machine learning model highly depends on the hyper parameter settings. Although effect of some parameters has been verified in the original paper of NGC (neural granger causality) [19], since this is the first study to use machine learning model for analyzing the whole-body human movement in a sports context within our best knowledge, we consider it is essential to sincerely provide how each hyperparameter affects the results and how we tuned those parameters for avoiding the overvaluation of our method. Specifically, in this supporting information file, we first describe the criteria for hyper parameter settings in this study, then, verified the relation between the changes of each hyper parameters and results of the analysis.

**Criteria for hyper parameter setting**

For setting the hyper parameters, variable usage rate, which indicates the ratio of candidate combinations of different body joints showing a positive NGC, was used for tuning of them. Figure S1 shows the effect of different variable usage rate on the causal graph obtained from the NGC analysis. If the value of variable usage rate is 0, there is no NGC among body joints and all joints' movement depends on the past information of themselves (i.e., autoregressive model). Also, if it shows 1.0, all joints have NGC to all other joints. Obviously, these two cases should be avoided. Specifically, we tuned the hyper parameters so that the mean variable usage rate among 16 pairs becomes around 0.25-0.50 based on the following insights. First, if the variable usage rate is over 0.50, there always inter-body joint combinations that have positive NGC, and this may lead the overestimation of the number and size of inter-body NGC. Also, to avoid the underestimation of the number of joint combinations that have NGC, we set the lower limit at 0.25 to capture about 50% of the NGC between inter-body joints (0.25/2+0.25/2). If there is any body joint that has large NGC to opponent's joints, we can also identify through this criterion. As a result of parameter tuning, mean variable usage rate among 16 pairs is $0.30 \pm 0.03$ in this study.

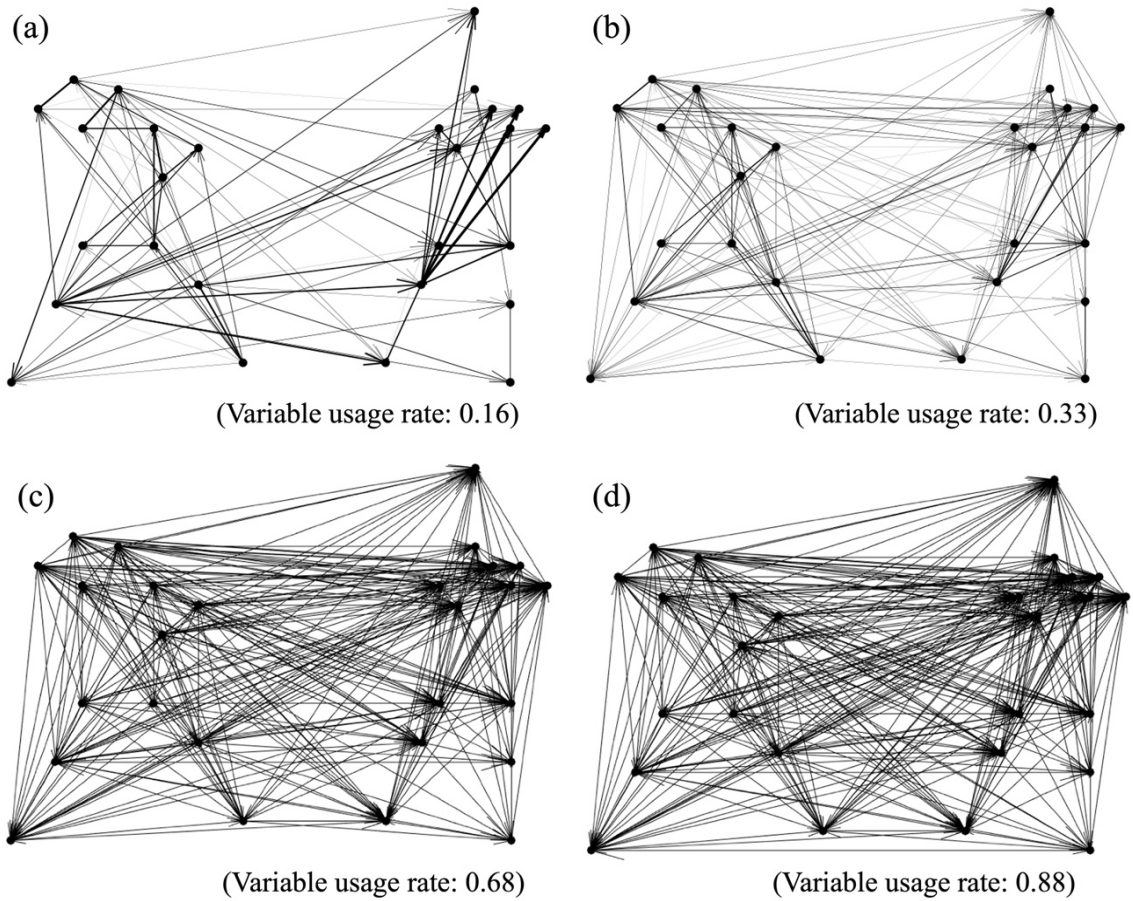

Figure S1. The relations between differences variable usage rate and obtained causal graph from NGC analysis (a) 0.16, (b) 0.33, (c), 0.68, (d) 0.88.

**Testing effect of hyper parameters on the result**

In this section, we verify the effect of hyper parameters on the result of the analysis of the paper. Specifically, we verify the relations between the changes of hyper parameters and variable usage rate, and the ration among four causal indexes ($ngc_{pp}$, $ngc_{bb}$, $ngc_{pb}$, $ngc_{bp}$) of the Analysis1 of the paper. Since the numerical value of NGC itself is not meaningful, we converted it to a ratio and examined how their relative size changes with parameter setting. The following are explanations of each parameter.

*penalty term $\lambda$*: this parameter penalizes the total amount of weights included in the trained model in equation (3). If we set smaller $\lambda$ value, the variable usage rate will be larger. We tested the impact of this parameter with different values of 0.0001, 0.001, 0.003 (our setting), 0.005, 0.01, 0.1 while keeping other parameters constant.

*Sampling rate (down sampling rate)*: this parameter controls the time interval between the points included in the input time series data. Since penalty term in equation (3) penalizes the sum of weights for avoiding to include too many lagged causalities, if we try to capture the characteristics of the IC with long time scale (several hundred milliseconds ~ several seconds), we need to down sample the data from the original motion capture data with high frequency. We tested the impact of this parameter with different values of 250 Hz (1.0), 125 Hz (2.0), 50 Hz (5.0, our setting), 25 Hz (10.0).

*Maximum lag*: this parameter controls the maximum time lag that is considered in the calculation. If there is any hypothesis about this parameter (e.g., near the ball travel time 0.50 sec for the baseball), it should be set as enough large to include it. We tested the impact of this parameter with different values of 2.0 sec, 1.0 sec (our setting), 0.5 sec, 0.1 sec. Note that although we only change the value of a single parameter in this file, the sampling rate should also be adjusted to keep the range of maximum time-lag const (e.g., 100 Hz with 50 points (0.50 sec) and 200 Hz with 100 points (0.50 sec)).

*Number of hidden units*: this parameter indicates the number of hidden units included in the layer of multi-layer perceptron. While the number of features that can be considered in the model increases as the number of hidden units increases, the complexly of the model and required time for the training process also increases. We tested the impact of this parameter with different values of 128, 64, 32 (our setting), 16, 8.

*Learning rate*: this parameter controls the step size for updating the weight parameters at each iteration based on the value of loss function. While if too small value of it increases the training time, too large value makes it difficult to find the optimal solution. We tested the impact of this parameter with different values of 0.1, 0.07, 0.05 (our setting), 0.03, 0.01.

*Type of input variable*: this parameter indicates the type of input variable to the model. Although this study tried to capture the characteristics of the interaction between two players from the aspects of the "velocity interaction", the form of the interaction may change depends on the type. Therefore, we also calculated the resultant acceleration data of each joint and verified this viewpoint.

**Results and discussions**

Table S1-S6 shows the effect of changing the hyper parameter on the mean value of variable usage rate and the ratio of four causal indexes ($ngc_{pp}$, $ngc_{bb}$, $ngc_{pb}$, $ngc_{bp}$) to sum of them. First, as shown in the Table S1, the size of $\lambda$ parameter affects the variable usage rate, i.e., if the penalty for the weight parameter increases, the number of causal relations detected through the NGC analysis will decrease. However, they are consistent with respect to their relative ratios to the sum of the weight parameters. Hence, the change of the variable usage rate could be due to an increase in the detection

of small causalities. Therefore, if criteria are set to eliminate combinations of joints with such a small causality, as in this study, we believe that a robust analysis is possible with respect to the increase or decrease of this parameter except for the case that the variable usage rate is 1.0 or 0.0. On the contrary, the size of the sampling rate affects the ratio of four causal indexes, i.e., the ratio of inter-personal NGC ($ngc_{pb}$ and $ngc_{bp}$) decreases as this parameter increases (Table S2). As mentioned above, this seems to be caused by the difficulty for detecting the long-time scale interaction (such as 0.50 sec delay) with high sampling rate. This is evidenced by the fact that the same trend is observed for the results of maximum time lag, i.e., if we allow only few time-lag, the inter-personal NGC decreases.

Further, from the results of number of hidden units (Table S4), this parameter has small effect on the analysis. The slight difference of the variable usage rate is thought to be caused by differences in convergence speed, i.e., larger number of hidden units require larger training time. However, if the model tries to identify the causal structure in more complex interaction, this parameter should be carefully tuned. Also, the learning rate has similar effect of the $\lambda$ parameter, i.e., it has large effect on the variable usage rate but not on the ratio of causal indexes (Table S5). The reason for increasement of variable usage rate in 0.1 is that it cannot converge to the optimal solution due to the large step size. Finally, slightly different results were observed by using resultant acceleration as the input data (Table S6). Namely, by using the acceleration data, the ratio of NGC among pitcher's body joints increased while that of batter's body joints decreased. Therefore, the type of input data may have some influence on the results even if the overall trend is unchanged. To account for the effects of amplified noise in acceleration data, this study adopts resultant velocity data as the input variable.

In summary of these results, although there are some sensitivities to the hyper parameter settings, we consider that the criteria and analytical procedures defined in this study have yielded robust results. Therefore, we believe that the setting of this study has a certain level of validity. However, when it applies to other type of interaction or human movement, the sensitive hyper parameters such as the down-sampling rate, maximum lag and type of input data should be carefully tuned.

Table S1. The impact of changing the value of $\lambda$ parameter on the results. NGC_pp, NGC_bb, NGC_pb, NGC_bp indicates the ratio of four causal indexes canulated in Analysis1 to the some of them.

| $\lambda$ value | Variable usage rate | NGC_pp | NGC_bb | NGC_pb | NGC_bp |
|---|---|---|---|---|---|
| 0.0001 | 1.00 ± 0.00 | 0.25 ± 0.00 | 0.25 ± 0.00 | 0.25 ± 0.00 | 0.25 ± 0.00 |
| 0.001 | 0.69 ± 0.03 | 0.47 ± 0.01 | 0.30 ± 0.10 | 0.21 ± 0.09 | 0.02 ± 0.01 |
| 0.003 (ours) | 0.30 ± 0.03 | 0.46 ± 0.03 | 0.33 ± 0.09 | 0.20 ± 0.09 | 0.01 ± 0.01 |
| 0.005 | 0.21 ± 0.02 | 0.44 ± 0.04 | 0.36 ± 0.09 | 0.19 ± 0.09 | 0.01 ± 0.01 |
| 0.01 | 0.15 ± 0.01 | 0.43 ± 0.05 | 0.36 ± 0.09 | 0.20 ± 0.10 | 0.01 ± 0.01 |
| 0.1 | 0.00 ± 0.00 | — | — | — | — |

Table S2. The impact of changing the value of sampling rate (down sampling rate) parameter on the results.

| Sampling rate (Hz) | Variable usage rate | NGC_pp | NGC_bb | NGC_pb | NGC_bp |
|---|---|---|---|---|---|
| 250 | 0.15 ± 0.01 | 0.41 ± 0.05 | 0.55 ± 0.05 | 0.03 ± 0.01 | 0.00 ± 0.00 |
| 125 | 0.19 ± 0.01 | 0.43 ± 0.04 | 0.48 ± 0.05 | 0.08 ± 0.03 | 0.01 ± 0.01 |
| 50 (ours) | 0.30 ± 0.03 | 0.46 ± 0.03 | 0.33 ± 0.09 | 0.20 ± 0.09 | 0.01 ± 0.01 |
| 25 | 0.33 ± 0.01 | 0.36 ± 0.16 | 0.33 ± 0.09 | 0.30 ± 0.16 | 0.00 ± 0.00 |

Table S3. The impact of changing the value of maximum lag parameter on the results.

| Maximum lag (sec) | Variable usage rate | NGC_pp | NGC_bb | NGC_pb | NGC_bp |
|---|---|---|---|---|---|
| 2.0 | 0.51 ± 0.05 | 0.30 ± 0.02 | 0.39 ± 0.15 | 0.29 ± 0.15 | 0.01 ± 0.01 |
| 1.0 (ours) | 0.30 ± 0.03 | 0.46 ± 0.03 | 0.33 ± 0.09 | 0.20 ± 0.09 | 0.01 ± 0.01 |
| 0.5 | 0.27 ± 0.02 | 0.45 ± 0.04 | 0.36 ± 0.07 | 0.17 ± 0.07 | 0.01 ± 0.01 |
| 0.1 | 0.25 ± 0.01 | 0.47 ± 0.03 | 0.48 ± 0.03 | 0.03 ± 0.01 | 0.01 ± 0.01 |

Table S4. The impact of changing the number of hidden units of the multi-layer perceptron on the results.

| Number of hidden units | Variable usage rate | NGC_pp | NGC_bb | NGC_pb | NGC_bp |
|---|---|---|---|---|---|
| 128 | 0.35 ± 0.04 | 0.46 ± 0.03 | 0.33 ± 0.09 | 0.20 ± 0.09 | 0.01 ± 0.01 |
| 64 | 0.31 ± 0.03 | 0.46 ± 0.04 | 0.33 ± 0.09 | 0.20 ± 0.09 | 0.01 ± 0.01 |
| 32 (ours) | 0.30 ± 0.03 | 0.46 ± 0.03 | 0.33 ± 0.09 | 0.20 ± 0.09 | 0.01 ± 0.01 |
| 16 | 0.29 ± 0.03 | 0.46 ± 0.04 | 0.34 ± 0.09 | 0.20 ± 0.09 | 0.01 ± 0.01 |
| 8 | 0.28 ± 0.03 | 0.46 ± 0.04 | 0.34 ± 0.09 | 0.19 ± 0.09 | 0.01 ± 0.01 |

Table S5. The impact of changing the value of learning rate parameter on the results.

| Learning rate | Variable usage rate | NGC_pp | NGC_bb | NGC_pb | NGC_bp |
|---|---|---|---|---|---|
| 0.1 | 0.50 ± 0.10 | 0.44 ± 0.04 | 0.35 ± 0.09 | 0.19 ± 0.09 | 0.01 ± 0.01 |
| 0.07 | 0.21 ± 0.02 | 0.44 ± 0.04 | 0.34 ± 0.09 | 0.21 ± 0.09 | 0.01 ± 0.01 |
| 0.05 (ours) | 0.30 ± 0.03 | 0.46 ± 0.03 | 0.33 ± 0.09 | 0.20 ± 0.09 | 0.01 ± 0.01 |
| 0.03 | 0.40 ± 0.04 | 0.46 ± 0.03 | 0.31 ± 0.10 | 0.21 ± 0.10 | 0.01 ± 0.02 |
| 0.01 | 1.00 ± 0.00 | 0.26 ± 0.00 | 0.25 ± 0.00 | 0.25 ± 0.00 | 0.24 ± 0.00 |

Table S6. The impact of changing the type of input variable on the results.

| Type of input variable | Variable usage rate | NGC_pp | NGC_bb | NGC_pb | NGC_bp |
|---|---|---|---|---|---|
| Resultant velocity (ours) | 0.30 ± 0.03 | 0.46 ± 0.03 | 0.33 ± 0.09 | 0.20 ± 0.09 | 0.01 ± 0.01 |
| Resultant acceleration | 0.52 ± 0.10 | 0.55 ± 0.07 | 0.22 ± 0.11 | 0.22 ± 0.11 | 0.01 ± 0.01 |